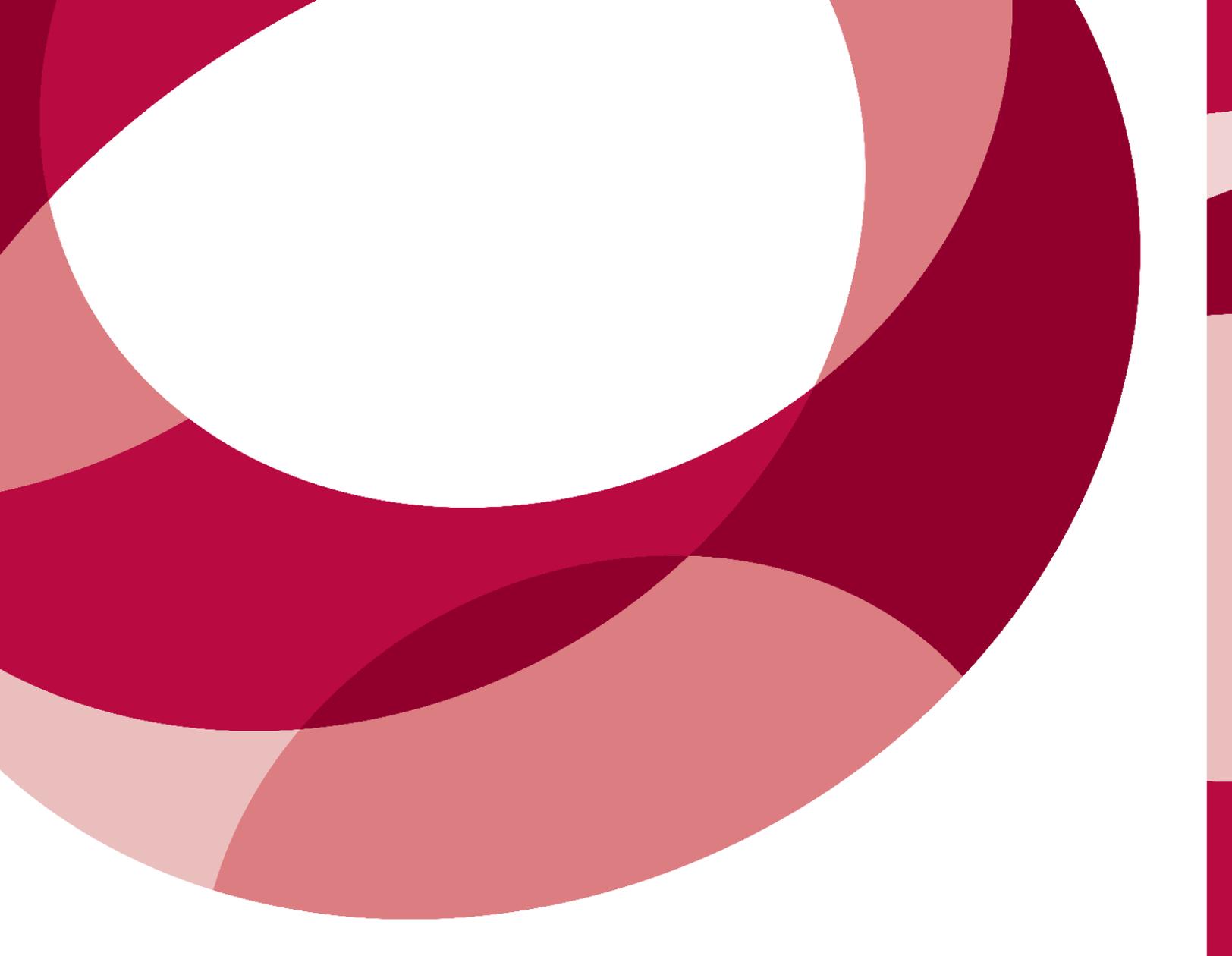

# Future of Information Retrieval Research in the Age of Generative AI

CCC Workshop Report

**December 2024**

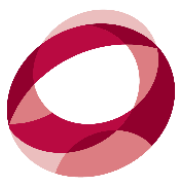

## Authors:

James Allan
University of Massachusetts Amherst, allan@cs.umass.edu, https://cs.umass.edu/~allan/

Eunsol Choi
University of Texas at Austin / New York University, eunsol@nyu.edu, https://eunsol.github.io/

Daniel P. Lopresti
Lehigh University / CCC, lopresti@cse.lehigh.edu, https://www.cse.lehigh.edu/~lopresti/

Hamed Zamani
University of Massachusetts Amherst, zamani@cs.umass.edu, https://groups.cs.umass.edu/zamani/

## With Support From:

Mary Lou Maher
Director of Research Community Initiatives, CCC, Computing Research Association
mmaher@cra.org

Haley Griffin
Senior Program Associate, CCC, Computing Research Association
hgriffin@cra.org

**Suggested Citation:**
Allan, J., Choi, E., Lopresti, D., & Zamani, H. (2024). Future of Information Retrieval Research in the Age of Generative AI CCC Workshop Report. Washington, D.C.: Computing Research Association (CRA). https://cra.org/wp-content/uploads/2024/12/Future-of-Information-Retrieval-Research-in-the-Age-of-Generative-AI.pdf.

# Table of Contents

# EXECUTIVE SUMMARY

In the fast-evolving field of information retrieval (IR), the integration of generative AI technologies such as large language models (LLMs) is transforming how users search for and interact with information. Recognizing this paradigm shift at the intersection of IR and generative AI (IR-GenAI), a visioning workshop supported by the Computing Community Consortium (CCC) was held in July 2024 to discuss the future of IR in the age of generative AI. This workshop convened 44 experts in information retrieval, natural language processing, human-computer interaction, and artificial intelligence from academia, industry, and government to explore how generative AI can enhance IR and vice versa, and to identify the major challenges and opportunities in this rapidly advancing field.

**Workshop Activities:** The workshop began with an overview of previous discussions from a preliminary brainstorming session and other related workshops and included presentations from selected participants to ignite deeper discussions. To explore specific themes, eight breakout sessions were then formed to discuss various aspects of the topic. The workshop agenda can be found on the CCC Website[1].

**Mechanisms for Effective Collaboration:** In order to effectively collaborate on the discussion topics, we followed the IDEO's brainstorming method (*7 Simple Rules of Brainstorming,* n.d.) based on the following seven mechanisms for effective collaboration: (1) defer judgments, (2) encourage wild ideas, (3) build on others' ideas, (4) stay focused on topic, (5) one conversation at a time, (6) be visual when needed, and (7) discuss as many ideas as possible (focus on quantity of ideas).

**Research Directions:** This report outlines eight research directions for IR-GenAI systems with high intellectual merits and broader impact: (1) Evaluation challenges and needs in IR-GenAI; (2) Learning from implicit and explicit human feedback for solving complex problems that may require reasoning; (3) Understanding and modeling users for the evolving generative AI-powered information access systems; (4) Challenges and potential solutions to address or mitigate socio-technical issues raised by the new technologies in IR-GenAI; (5) Methods for developing personalized IR-GenAI systems; (6) Efficiency considerations when scaling compute, data, and human efforts in developing IR-GenAI methods; (7) The role of information retrieval in enhancing AI agents; and (8) Developing foundation models specifically for information access and discovery.

This report contains a summary of discussions as potentially important research topics and contains a list of *recommendations* for academics, industry practitioners, institutions, evaluation campaigns, and funding agencies.

---

[1] https://cra.org/ccc/events/future-of-information-retrieval-research-in-the-age-of-generative-ai-ccc-visioning-workshop/#agenda

# 1. INTRODUCTION

## 1.1 Background and workshop goals

In today's rapidly evolving digital landscape, the field of information retrieval (IR) is at the intersection of traditional search methodologies and cutting-edge machine learning and artificial intelligence (AI) technologies. As we witness the proliferation of generative AI-driven models, such as diffusion and large language models (LLMs), it has become increasingly evident that the boundaries of IR research are expanding. The ways users search for information and the ways systems recommend information to users are already being impacted by generative AI technologies. Conversely, information retrieval technologies can have substantial impact on generative AI applications in terms of efficiency, effectiveness, robustness, and trustworthiness. Retrieval-augmented language models are just one example of direct impact that has recently attracted considerable attention in both academia and industry.[2] A few of these areas have been discussed in two visioning perspective papers on "Retrieval-Enhanced Machine Learning" (Zamani et al., 2022) and "Large Language Models and Future of Information Retrieval" (Zhai, 2024) and multiple workshops including "Search Futures" (Azzopardi et al, 2024) and "Task Focused IR in the Era of Generative AI" (Shah & White, 2024). These areas are becoming increasingly important every day. This growth of activity is why we believe this is the right time to discuss major challenges and opportunities in shaping the *Future of Information Retrieval Research in the Age of Generative AI.* This has motivated us to organize a visioning workshop on this topic with the support of the Computing Community Consortium (CCC).

To explore and navigate this dynamic landscape, we gathered a group of 44 experts across academia, industry, and government in the fields of information retrieval, natural language processing, human-computer interaction, machine learning, and broadly artificial intelligence to Washington, D.C. for this visioning workshop.[3] The group came together to outline the future of IR research, with generative AI playing a central role in reshaping how we discover and interact with information. This report is a reflection of the themes, recommendations, and ideas that emerged during the visioning workshop. The workshop used the Chatham House rule,[4] so the ideas are not attributed to specific participants. This report is a synthesis of the discussions during the workshop and does not include citations as in a traditional journal article.

From the visioning workshop discussions, two major themes emerged: (1) enhancing generative AI models and applications using information retrieval techniques, and (2) enhancing information retrieval models and applications using generative AI techniques. We acknowledge that it is sometimes difficult to draw the boundaries between IR and generative AI systems and these two systems are intertwined from many aspects. The ill-defined boundary between the two areas was brought up and thoroughly discussed during the workshop and the participants acknowledge and stress that information retrieval is not (and never has been) limited to current search engine systems, but broadly addresses how users or machines look for, find, present, access, discover, and interact with information. This creates a large number of challenges that information retrieval and other AI communities urgently need to tackle in the next five to ten years, and corresponding research opportunities that are available. Important real-world applications that can benefit from research in this area include, but are not limited to, search engines, recommender systems, question answering systems, dialogue systems, and intelligent assistants.

---

[2] In the last couple of years, Google Scholar indexed over 6,000 papers related to the phrase "retrieval-augmented generation" (as of July 2024). Although we have not counted the number of citations those papers generated, the impact of this area is clearly substantial.
[3] See Appendix A.2 for the complete list of workshop participants.
[4] https://www.chathamhouse.org/about-us/chatham-house-rule

We want to stress that this workshop focuses on challenges and opportunities that arise at the *intersection* of IR and generative AI fields and is not intended to cover the union of the two fields. There are additional issues that arise within generative AI independent of IR and within IR that are unrelated to generative AI. This report makes no attempt to discuss those issues.

## 2. HOW THIS DOCUMENT CAME ABOUT

### 2.1. Pre-workshop activities: How we assembled

Given the importance of the topic, the UMass Amherst Center for Intelligent Information Retrieval (CIIR) organized a relatively small, regional, invitation-only *Brainstorming Session on Information Retrieval Research in the Age of Generative AI* on December 1-2, 2023, in Amherst, Massachusetts (*CIIR hosts Two-Day Brainstorming Session on IR Research in the Age of Generative AI*, n.d.). The brainstorming session included seven faculty members from University of Massachusetts Amherst (UMass), six faculty members or senior researchers from five other institutions, and 15 doctoral students from UMass and Carnegie Mellon University. In this two-day event, the future challenges in the intersection of IR and generative AI research were discussed and there was a unanimous consensus that the organization of a follow-up workshop was needed.

The organizers thus reached out to the CCC with a proposal to run this visioning workshop. Upon CCC's approval, the organizers worked together to expand the organization team and compile a list of potential participants from diverse research communities, backgrounds, demographics, and institutions. A questionnaire was sent to the potential participants to collect feedback on their interest in joining the visioning workshop and preference on timing and location. Given the feedback from over 30 potential participants, the organizers and the CCC decided to hold the visioning workshop in Washington DC on July 19-20, 2024, immediately following the ACM SIGIR International Conference on Research and Development in Information Retrieval, the premier conference for information retrieval research.

A few weeks before the workshop, the participants were asked via a second survey to think about the core topics of the workshop and provide their individual ideas and indicate which of those particularly interested them. The responses were used by the organizers to frame initial ideas for the workshop.

### 2.2. Workshop activities: What we discussed

In all, 44 experts from various computing disciplines, including 40 invited researchers and 4 organizers, in addition to 5 members of the CRA staff and 1 member of the CIIR staff joined the visioning workshop in person at the Planet Word Museum in Washington, D.C. The participating experts comprised 30 academics, 5 government employees, and 9 industrial researchers; 34 US-based attendees and 10 from outside the US; 24 primarily affiliated with information retrieval, 12 with NLP, and 8 with other AI topics. A list of participants and their affiliations is provided in the appendix.

The workshop was structured as follows:

- **Opening:** The organizers kicked off the workshop by describing the visioning process, defining the scope of the IR-GenAI workshop, highlighting participation rules for breakout sessions, and presenting the workshop agenda. There also was a quick introduction of each participant.

- **Kicking off the discussions:** The organizers presented an overview of the discussions that happened during the CIIR Brainstorming Session on IR Research in the Age of Generative AI and the MSR Workshop on Task Focused IR in the Era of Generate AI. Seven participants were selected by the organizers based on their responses to the pre-workshop questionnaire, aiming for a diverse set of topics of broad interest. They each presented a two-minute visioning idea (a "flash prompt") to catalyze discussions on those and related topics. These talks were followed by substantial open-floor discussions suggesting alternative topics, expanding on those previously offered, and offering combinations and divisions. Workshop organizers, staff, and participants synchronously took notes on the discussions in a shared document and added additional possible topics or thoughts within the document.

- **Forming breakout sessions:** The participants then compiled a list of potential breakout sessions. (We also used existing commercial generative AI technologies to produce a list of breakout sessions from the notes; unfortunately without useful output.) The list of potential breakout sessions was iteratively refined based on the feedback from the participants in an open-floor discussion and periodic votes of interest, resulting in eight breakout sessions within IR-GenAI:

    a. **Evaluation:** This breakout group focused on (1) developing datasets, methods, and platforms for evaluating the next generation of (LLM-powered) information access systems, and (2) using generative AI technologies to develop more reliable and cheaper (personalized) evaluation methodologies for existing information access tasks.

    b. **Training, Feedback, and Reasoning:** This breakout group discussed (1) effective methods to collect and use explicit or implicit feedback from users for training the next generation of retrieval-enhanced generative AI systems, and (2) the obstacles and potential solutions in unlocking higher-level reasoning capabilities in the current generative AI technologies.

    c. **Understanding and Modeling Users:** This group discussed major research questions related to what users need and expect from new generative AI-powered information access systems, how to build effective user models for these new technologies, and how to mitigate issues related to ethics and privacy.

    d. **Social Ramifications:** This breakout group focused on socio-technical challenges in developing the new generation of information access systems and potential solutions to mitigate or address societal and ethical issues these technologies may create.

e. **Personalization:** This breakout group explored challenges, opportunities, and potential solutions in developing "digital twins" or "digital shadows" for users to develop effective personal assistants for information discovery and access. These terms are defined in Appendix A.1.2.

f. **Scaling Across Compute, Data, and Human Efforts:** This group built on the idea that current success in deep learning and generative AI research is largely due to various scaling efforts, to discuss potential challenges and opportunities in scaling efforts and how to continue this progress in the scope of IR and generative AI research efficiently.

g. **AI Agents and Information Retrieval:** This group discussed the challenges and opportunities in developing intelligent agents that are ubiquitous, effective, inexpensive and able to provide information, or accomplish tasks on behalf of users. This group also explored how different agents can interactively communicate and solve complex problems, such as planning and decision making.

h. **Foundation Models for Information Access and Discovery:** This breakout group focused on the needs for and potential next steps in developing foundation models, specifically designed for information access and discovery.

- **Breakout sessions:** Each participant chose one breakout session and participated in detailed discussions throughout Friday afternoon and Saturday. Each breakout group had chosen a leader and a scribe, though writing was typically distributed throughout the group. Each breakout session presented a brief oral summary of their work-so-far to all participants twice, with responses by the full set of participants entered directly into that group's notes. This process ensured that all voices were heard and feedback was received. The last part of the workshop was devoted to writing up descriptions of sessions and compiling the challenges and opportunities discussed in the breakout session (and presented below).

- **Closing:** Each breakout session presented two things to the entire set of participants: (1) a major and compelling recommendation that arose in their discussions and (2) a burning question whose answer could improve the group's report. The workshop ended with an open-floor discussion of the questions, challenges, and any other topic that someone felt needed to be raised.

## 2.3. Post-workshop activities: How we produced this report

The workshop organizers gathered immediately after the workshop, read the reports from each breakout session, and discussed the outline and potential content of the report. The first draft of the report was produced solely by the organizers using all the material produced by each breakout session (i.e., notes, summaries, comments, challenges, questions, recommendations) and extending or modifying them as needed. The draft of the report was shared with CCC and workshop participants for further feedback and it was revised to shape the report at hand. This report has been reviewed internally by a member of the CCC council and externally by 2 members of the IR/AI research community.

# 3. SUMMARY OF THE DISCUSSED RESEARCH TOPICS FOR FUTURE EXPLORATION

Here we present a brief summary of the key observations, challenges, and opportunities discussed in each breakout session. More details on each topic are provided in the following sections.

## 3.1. Evaluation

- Exploring the limits of using LLMs to label material as relevant or not, including automatically generated explanations that indicate *why* something is relevant (or not).

- Expanding evaluation approaches to include the entire search and discovery process: handling multi-step processes such as conversations, measuring the accuracy of generated (rather than simply retrieved) responses, and designing approaches that measure the interactions between IR and GenAI.

- Developing and employing "digital twin" technology to enable strong and reliable simulated user evaluations. That includes using variations of the same "twin" to more broadly understand the range of human responses.

## 3.2. Training, feedback, and reasoning

- Developing interactive generative AI and information retrieval systems that cooperatively learn when to submit queries and what information to retrieve that augments the knowledge encoded in generative AI systems.

- Exploring various implicit and explicit feedback mechanisms to iteratively improve retrieval-enhanced generative AI systems.

- Retrieving, organizing, and synthesizing vast amounts of information using generative AI systems that can manage complex reasoning tasks.

## 3.3. Understanding and modeling users

- Understanding user needs and their multimodal interactions with generative AI systems for information discovery and access.

- Learning an effective model of users by representing their cognitive state, including their state of knowledge, while using generative AI systems for information discovery and access.

- Exploring privacy-preserving solutions in modeling users in generative AI systems.

## 3.4. Social ramifications

- Thinking about the impact on all aspects of society continually: before, during, and after development.

- Identifying risks, challenges, and opportunities of generative AI for information retrieval requires an interdisciplinary approach informed by socio-technical perspectives and co-developed with social science scholars, legal scholars, civil society representatives, and policy makers among others.

### 3.5. Personalization

- Developing efficient, personalized generative AI systems that act as digital twins to learn personal behavior and preferences of the user for personalized retrieval, recommendation, and synthesis of information.

- Developing persuasive recommender systems that not only produce accurate recommendations but also provide sufficient explanation, transparency, and justification to persuade their users.

- Developing efficient personalized generative AI systems for on-device intelligent assistance and information access.

### 3.6. Reducing the cost of generative IR

- Exploring the capabilities of numerous small LLMs rather than a single monolithic approach.

- Investigating the tradeoffs between different sized LLMs and different sized IR systems.

- Exploring ways to accomplish the same (or improved) capabilities with a fixed hardware footprint rather than one that is assumed to grow continuously.

- Considering the impact of new computational paradigms (e.g., quantum, bio, or neuromorphic computing) to understand their impact on GenAI and IR.

### 3.7. AI agents and information retrieval

- Understanding how mixed-initiative systems leverage GenAI and IR to gather information proactively for users without direct initiation.

- Exploring how agents can communicate with each other reliably, particularly in the face of current hallucination challenges.

- Developing approaches that allow GenAI to make novel and perhaps complex plans for gathering information in response to challenging questions.

- Creating evaluation frameworks for groups of agents, some of which address multimodal inputs, collaborating with each other and users to accomplish a task.

### 3.8. Foundation models for information access and discovery

- Developing a task-agnostic foundation model for information access, extending human intelligence with personalized, contextual, and multimodal information and beyond.

- Developing techniques for better incorporating user behavior, supporting multiple modalities, and generating accessible output.

- Extending foundation models so that they can continuously learn, improve, avoid or address biases, and so on.
- Creating non-profit regional organizations that aim to solicit computing resources to be shared among regional researchers.

# 4. SHORT- AND LONG-TERM RESEARCH TOPICS AND RECOMMENDATIONS

In this section we provide more context, additional details, and extra challenges and opportunities from each breakout group, building on the summaries provided above. Each includes summary information followed by a list of short-term (within five years) and long-term (within ten years) research challenges.

## 4.1. Evaluation

We identified two broad classes of evaluation challenges: (1) the application of genAI to IR evaluation and (2) the evaluation challenges of generative IR.

Generative AI offers two key opportunities to the evaluation of classic document or passage retrieval systems. First, a classic approach to offline evaluation in IR is the use of a test collection. The construction of these collections is resource intensive, in particular the creation of the lists of the documents that are relevant or not relevant to queries. There is growing evidence that LLMs may be useful in labeling documents. Research to explore and refine this labeling process holds great promise.

A second key aspect of offline evaluation is the so-called user evaluation: the observation of people in how they initiate or react to the actions of a search engine. There is the prospect of exploiting GenAI to simulate the actions of people. One might even speculate that human "digital twins" could be created for the purposes of testing an information retrieval system.

For the purposes of evaluating a generative IR system, there is much to consider. Retrieval Augmented Generation represents a coalition of IR and generative systems to accomplish a task. It is necessary to consider evaluation of these component systems in concert as well as individually. It is also necessary to establish the domain of competency of generative IR systems. A challenge with the output of a generative system is that it always expresses an answer with apparent confidence. There is great value in knowing the knowledge domain that a system is competent in. How do we establish and communicate that competency? In a related topic, we don't know how an LLM is achieving its answers. A research challenge is to establish a methodology for building confidence in the output of LLMs. Finally the reproducibility of systems needs to be considered.

Note that recommendations for carrying out community-wide evaluations – some of which might be impacted by the research suggested here – are presented later in this report.

### 4.1.1. Recommendations for short-term (the next five years) research programs

*4.1.1.a. Using LLMs to support the relevance assessment/labeling process*
To support the evaluation of classic search, researchers rely on a fixed set of information needs and the identification of relevant documents. There is currently an explosion of interest in having LLMs do the labeling of documents to control costs. There is a spectrum of ways to use LLMs to that end, from replacing human judgments with LLM labels to utilizing them as an assistant. Using LLMs as labelers directly suffers from various issues. For instance, LLMs make mistakes, create misinformation through hallucination, and suffer from various types of biases. Alternatively one could use an LLM to assist an assessor who is the one responsible for providing the labels. In particular an LLM could help assessors be more consistent by having LLMs partner with assessors to identify documents that may need to be rejudged because relevance judgments are inconsistent. Here continual fine-tuning of a personal, topic-specific LLM could support label suggestion.

*4.1.1.b. Proposing relevance rationales (one-line comments) and highlighting*
For many evaluation data sets, relevance labels are applied at the document level, but there is no recording of why a document was labeled as relevant or – when not all of the document is relevant to the information need – which part of the document contains the relevant information. A research direction could investigate whether LLMs could be used to generate written descriptions of why a particular document is relevant. Alternatively, LLMs could be used to identify which particular parts of a document provide relevant information.

*4.1.1.c. Evaluation protocols able to assess the RAG pipeline as a whole, digging into the interaction between retrieval and generative components*
The current evaluation methodologies for Retrieval-Augmented Generation (RAG) pipelines assess retrieval and generation separately, assuming that the better the retrieval (evaluated on its own), the better will the generation (evaluated on its own) based on the top-ranked documents. However, there is not an a priori guarantee that the text of the most relevant document used as a prompt will lead to the generation of the best output; it could be that the second or the third relevant document would provide a better priming of the LLM, inducing it to produce a better response. Current evaluation is not designed to investigate these aspects and this kind of interaction and there is an opportunity to develop more comprehensive approaches able to assess the whole RAG pipeline and clarify the effects of this interaction. The benefit of these new evaluation methods would be to not only lead to improving RAG itself but also to shed some light on the inner workings of LLMs.

*4.1.1.d. Domain of competency for LLMs and Prompt Performance Prediction*
It is currently thought that LLMs embed knowledge distilled through their learning process, but it is not fully clear what this knowledge actually is and what domains it covers. There is a broad understanding that general purpose models need to be specialized or fine-tuned to specific domains and this led to a proliferation of models for specific languages or areas. There is also the idea that bigger and bigger models should be able to perform better and better in more and more domains, even if it could still be the case that they lack enough training data in a given domain or that the relative size of the training data for a domain becomes smaller and smaller as the other domains grow bigger, making the model less effective in that domain. In general, there is a lack of an evaluation methodology able to investigate and determine which is the actual domain of competence of an LLM and/or how an LLM could perform in a given domain (or outside it). This kind of evaluation would be needed for understanding the extent to which an LLM is suitable for one or more domains. Moreover, it would open the possibility for a new area of investigation, i.e. Prompt Performance Prediction (PPP), aimed at estimating whether a given prompt could be effectively dealt with by an LLM, given its domain(s) of competency.

*4.1.1.e. Guarantee reproducibility and generalizability of the experiments, in order to improve the validity of the conclusions drawn, e.g. by controlling for the effect of (small) variations in prompts*

Reproducibility is a primary concern in almost any area of science and it has been discussed in IR for many years as well. LLMs, and particularly proprietary LLMs, bring the reproducibility concerns to a completely new and different level. There is an urgent need for much more "open source" and shared models and datasets, in terms of code, training data, training regimes, intermediate snapshots, and so on. This is needed just to be able to conduct experiments on a common ground and to strive for reproducibility. Another angle of reproducibility concerns the extreme variability in the functioning of LLMs themselves where even small changes in the prompt may lead to substantially different answers. Therefore, there is a need for a more systematic investigation of the behavior of such systems in order to understand how to control the effect of such small variations or how to account for them when assessing and comparing systems, e.g. by suitably computed "confidence intervals". The ultimate goal is to improve the validity and generalizability of the conclusions drawn from the experiments.

### 4.1.2. Recommendations for long-term (the next ten years) research programs

*4.1.2.a. Simulating user studies and A/B tests / Digital Twins*

Simulation has been pursued for a long time in IR, and LLMs provide a concrete opportunity to push its boundaries and open up to the evaluating system by credibly simulating the user behavior or even by creating whole synthetic experimental collections (e.g., topics, documents, relevance judgements, clicks). Ultimately this kind of simulation would correspond to the creation of *digital twins* of people/users (or better of groups of people/users) which can be exploited to assess systems. Such digital twins could be useful when testing new systems for which there are no existing interactions and one wants to be proactive. Such a simulation would clearly offer the possibility to scale-up and make evaluation much more fine-grained, allowing for more systematic testing before moving to the user-side. Simulation naturally fits in the continuum of evaluation which ranges from lab-based evaluation based on experimental collections, i.e. a kind of static simulation of user information needs and answers to them, to interactive evaluation, e.g. A/B testing with real users in an operational setting. One open question is whether current LLMs are enough to create such digital twins or whether we should imagine more advanced models which, besides the language generation capability, include behavioral models of users, browsing or interaction components, and more. In some sense, these enhanced models could parallel what multimedia models are today, instead of being constituted by multiple media, they would be constituted by multiple components of such a digital twin. The perspective of such models raises several questions: how do we validate them? How stable would they be? Would their training data be rich and diverse enough to (even partially) capture what a real person is? It is important to recall that this will also be an approximation of users and thus in the end cannot substitute for actual user studies. Note that developing "digital twins" is also discussed in Section 4.5 for a different purpose, which is related to user simulation but has some fundamental differences.

*4.1.2.b. Designing evaluation that is capable of accumulating evidence about an LLM's ability to provide the correct answer because of the correct processing/inference*

LLMs are systems initially designed for a specific task, i.e. language generation, that are now used for many different purposes and downstream tasks. The current evaluation methodology is designed to assess systems that were specifically designed for one single purpose, e.g., sentiment classification. As a consequence, it is based on verifying that for a given input, the expected output is produced, e.g., positive or negative sentiment. However, LLMs are not designed for a specific purpose and there is not an a priori guarantee that the correct answer is generated because of the correct process/inference/reasoning. There is a need and an opportunity for developing deeper evaluation methodologies to investigate and provide evidence about how and why a given answer has been generated. One can consider this linked to some sort of "explainability", or can view this as a kind of "debugging", or if one imagines an LLM as a big matrix or an extremely complex circuit, this kind of evaluation tries to understand which parts of the matrix (or the circuit) deliver classification, which ones ranking, etc. In this sense, this is different from AI explainability which is instead targeted at providing an explanation for a specific answer, e.g. which features mattered most. We note that this is related to the "Domain of Competency" listed as a short-term challenge.

*4.1.2.c. Evaluating results of systems that generate responses rather than rank documents*

Responses to information needs that are knowledge requests where a generative response is appropriate will need new approaches to evaluation. This is in contrast to item-based retrieval where a ranked list is an appropriate response. The overarching evaluation of generated output will need to verify that the output is responsive to the information need. Different evaluation approaches will be needed depending on whether the generated output includes citations to source documents or not. With source documents, existing approaches that utilize document labels could be integrated into the evaluation; however, whether or not there are document citations, there will need to be approaches to evaluating whether or not the content that appears in the generated text is responsive to the request. LLMs have a role to play in answering such questions. It is likely that in addition to new frameworks for evaluation, new metrics for evaluation will be needed. The reproducibility of the evaluation is an important consideration and how to perform the evaluation in a way that enables consistent comparison of systems (ie. techniques) is also critical.

## 4.2. Training, feedback, and reasoning

At the heart of retrieval-augmented generation (RAG) systems, and more broadly retrieval-enhanced generative AI (RE-GenAI) systems, is the interaction between the information retrieval / access model and the language model (LM), and eventually the user. Current training methods focus on specific and narrow definition of RE-GenAI and mainly focus on training RAG components separately, but we see promise in tightly coupling the training of the information access and language model, because such training would allow components to adapt to one another for the purpose of doing well on the downstream task. Some near challenges in joint training are to learn when to retrieve (e.g., based on measures of model uncertainty), when to rely on parametric knowledge, and how to deal with conflicting information.

Moreover, the granularity and the representation space of retrieval are also unclear, i.e., retrieving sentences, passages, entire documents, or even latent representations from a memory. When it comes to new sources of human feedback, it is unclear what constitutes meaningful feedback when click data (user clicks on relevant documents) is no longer available. Some sources of implicit feedback include absence of follow-up questions or question reformulation, which provide explicit scalar or natural language feedback, and how to perform credit assignment from target answer to source documents. We also envision that components can self-improve – the information access and LM components can provide training signals to one another. Nevertheless, we realize the importance of human annotations, and advocate for a mix of human and machine annotations and argue research on principled integration of human and machine labeling is required. We also envision some paradigm changes where retrieval is possible even when it requires complex reasoning over document collections, e.g., when queries are not similar to the underlying knowledge but are more abstract and/or require aggregation and synthesis at different granularity.

### 4.2.1. Recommendations for short-term (the next five years) research programs

***4.2.1.a. Joint and end-to-end training of retrieval and language models at both pre-training and fine-tuning phases***

The information access models in RE-GenAI systems may benefit from being trained with awareness of the GenAI model. Similarly, the GenAI model may benefit from being trained with awareness of the information they are likely to encounter as the output of an information access system. This diverges from current practice where the information access and the GenAI systems are trained separately and only combined at test time. Even though there are some preliminary studies in this space, many open research questions exist, such as learning optimal communication protocols and languages between IR and GenAI systems. See *4.1.1.c* for recommendations for evaluating such a pipeline.

***4.2.1.b. Learning when to retrieve vs. generate***

Current RAG systems are typically implemented as a "hard pipeline": retrieve, then feed the inputs into the reader for generation of the answer. Some work has shown how to do retrieval more dynamically, either with sequential queries or even based on the output of previous outputs from information access systems. A key question here is how to estimate generation uncertainty: if we know when a language model will produce an inaccurate answer, we have found a point in the generation where retrieval is more likely to improve the reliability of the results. Defining optimal communication language between GenAI and retrieval models is of significant importance to make retrieval versus generation decisions.

***4.2.1.c. Uncertainty modeling in alignment***

Related to the above point, we expect it to be important to develop techniques for calibration of LLMs in the context of alignment. For language modeling, it is still not understood how different alignment techniques (RLHF, direct preference optimization (DPO)) impact calibration. For retrieval-augmented settings, the uncertainty estimation needs to consider not just the language model's behavior and output, but also the retrieved documents.

*Future of Information Retrieval Research in the Age of Generative AI*     15

*4.2.1.d. Self-improving RAG systems*

Systems where the retrieval and generation component provide training signals to one another to progressively and jointly improve (what are good search queries, what do good documents look like).

*4.2.1.e. Training better retrievers given feedback from users only on generated answers*

Unlike in search engines where users can provide explicit and implicit feedback at document level (e.g., by clicking on documents in the search engine result page), translating user feedback on generated answers to document-level feedback on retrieved information is challenging. Future work can potentially address these issues through: (1) credit assignment from generated answers to source documents. Can we identify where a generated answer referenced a particular source document? This potentially requires reader models that generate attributed answers or interpretability techniques that enable us to recognize and manipulate how closely the generated answer matches the source. (2) coming up with new training signals based on user behavior (e.g., eye tracking, session data, interactions with citations). See *4.1.2.c* for ideas about evaluation when there are generated answers.

*4.2.1.f. Training from a mix of human annotation and machine labeling*

We envision a hybrid paradigm where the retriever is trained with machine labels as above, or humans can directly intervene to indicate where poor retrieval has hampered the performance of the system. Again, this notion requires some notion of epistemic uncertainty to know where human annotation is necessary. This can be further improved by leveraging methods from active inference (i.e., choose examples to evaluate on rather than train on for reliable unbiased estimation of model performance).

*4.2.1.g. Training a retrieval model (a search engine) for multiple GenAI Systems*

Similar to search engines that provide service to many users and learn from their interactions, we can envision a retrieval model or a search engine that provides service to many GenAI systems, depending on their needs. These systems need to formalize a standard query language that is shared among different GenAI systems. A major benefit in working on this approach is that it can benefit from feedback from a diverse set of GenAI systems, helping the retrieval model to generalize across GenAI models and learn more effective retrieval functions. Calibrating feedback across GenAI systems in order to optimize the search engine is among the important research directions in this area. Personalizing the result list for each GenAI system is of great importance, as each demonstrates different behavior and consumes the result lists differently.

*4.2.1.h. Developing information access models that require reasoning*

Complex questions that are posed to GenAI systems are often different from the types of queries fed into conventional search engines. Queries may require additional reasoning, which poses problems for search engines. Queries may also require skills like aggregating over opinions, e.g., to summarize or give insights from product reviews. We believe that additional benchmarks and reliable, reusable, and scalable evaluation methodologies should be developed to test this setting. Furthermore, different types of retrieval machinery may be needed.

*4.2.1.i. Instruction-following information access models*

How can we get IR systems to follow prompts in queries similar to LLMs, which define constraints external to the semantics of queries. To develop such systems, we must (1) move away from symmetric encoders as the query encoders need to incorporate instructions in addition to query text, (2) define constraints on properties of the documents - prefer formal sources, prefer personal opinions, etc. (3) invest in developing generative information access models. There is no doubt that instruction-following retrieval models are necessary to build general-purpose retrieval models. Progress in this area requires large-scale training datasets that include a diverse set of instructions. Generalization to unseen instructions is a significant challenge in this area.

4.2.2. Recommendations for long-term (the next ten years) research programs

*4.2.2.a. Fresh generative models*

As opposed to using RAG to put information into the context of an LLM, we can imagine future systems where retrieved documents are directly used to update the parametric knowledge of an LLM on the fly. For instance, knowledge editing methods promise to enable updates that LLMs can successfully reproduce to answer new queries. Can we get to a point where a new document can be added to an LLM and reproduced losslessly when the LLM is queried appropriately? Challenges include the difficulty of knowledge editing methods to correctly update the desired knowledge.

*4.2.2.b. Continual joint training of parametric and nonparametric memory*

When models are trained from scratch, they encode some facts reliably in their parametric knowledge. As we develop RE-GenAI systems, we may wish to establish a boundary between what information needs to be memorized (and memorized completely) vs. what information a language model should ignore and assume it will be provided at test time. A model should learn what to put in each type of memory, and when to take non-parametric information and on-the-fly move it into parametric memory when it proves generally useful.

*4.2.2.c. Implicit user feedback modeling*

Current feedback channels include explicit supervision and implicit supervision from users like clicking on links in generated responses. However, generative AI systems bring up a host of new challenges and potential new feedback types given the emphasis on long-form text generation. For instance, users may re-query an LLM with a modified version of their query, like in traditional IR, but they also may simply ask the LLM "show me something different" or "elaborate on [aspect] of the [response]". Additionally, the way a user processes LLM responses could be further studied with mechanisms like gaze tracking. All of these contribute possible methods for additional user feedback.

*4.2.2.d. Reasoning over user context and recommendations*

The input to the retrieval model from the GenAI system does not have to be an explicit query describing the information need (in either natural language or latent representation). Instead the input could be some contextual information (such as user's short- or long-term history, user's location, device, time, etc.) that could initiate a recommendation functionality depending on the context (without an explicit query). This will enable us to develop systems that consume, synthesize and reason over the provided contextual information and recommended items and proactively interact with the users when needed. Such advancements are necessary to develop truly mixed-initiative intelligent agents.

## 4.3. Understanding and modeling users

Generative AI provides us with new capabilities that allow us to imagine richer interactions with users. Future information access systems and intelligent agents should offer richer understanding of the user's tasks, goals, context, experience, and cognitive state. To support complex tasks and decisions, these systems should generate plans to decompose the tasks and track task progress and user state jointly. The systems should be more interactive and proactive, offering the opportunities to users to do multi-turn interactions, to integrate multimodal data, and to accomplish more complex tasks. The new interactive interface raises new challenges that should be dealt with in future research: user goal detection, integration of rich context information, task and user modeling, planning and choosing appropriate actions and determining the right type of answer or results for the user. GenAI can help implement these functionalities but we will also need to develop specific models for them. In addition, data and evaluation will be important challenges. What is envisioned for future users is an integrated system assisting users to accomplish various tasks that often require information access.

### 4.3.1. Recommendations for short-term (the next five years) research programs

*4.3.1.a. Multimodal interactions*
In the next five years, we expect that the GenAI research community will make advances in foundation models, creating the ability to query and return results from multimodal data. The research community could look to make progress toward intent detection in multimodal context: what is the user's need in the current context?

*4.3.1.b. New information access use cases enabled by generative AI*
The current chatbot model focuses on linear question-response multi-turn interactions, with some customization based on prior statements or conversations. However, this ad hoc query approach encompasses only one of many potential use cases for GenAI. A gap that research could address is the creation of a taxonomy of prompts in the context of GenAI. A unified set of terminology could aid the research space in identifying new use cases for LLMs and other GenAI, including push-based models that provide information to users without the need for a stated query. Based on the identified needs, a system should identify the right type of input (multimodal) and response—including problem breakdown and generation of single mode or multimodal responses. Evaluation approaches (as described elsewhere in this report) could focus on the extent to which users' diverse needs are met.

*4.3.1.c. Developing robust user models within IR-GenAI systems*
An important prerequisite for achieving the aforementioned future goals is a more robust model of the user. At present, there are two disparate representations of the user: the model that the user would use to describe themselves and their interests, and the representation that a system has for the user. In the future, we would expect to bring these closer together–creating a cognitive state representation of the user, which could be operationalized by LLMs and other GenAI. Given the need to maintain accuracy, the cognitive model should be both explainable to the user, as well as correctable by the user—potentially supported by natural language conversation with the model of the user. Additionally, there is a need to evaluate the representativeness of the model of the user.

### *4.3.1.d. Privacy and economic implications of IR-GenAI technology*

To what extent is this model an accurate representation of the user? Taking a cue from eDiscovery and posthumous search spaces (i.e., selective disclosure), identifying what information should be provided or withheld from systems for privacy reasons could be treated as an information retrieval problem. Given the potential level of detail provided by and about the user, should they be able to control sharing of this data–for instance, by receiving payment for their personal model?

### *4.3.1.e. Cultural, social biases, and trustworthiness in user models*

Over the next few years, strides should be made toward modeling biases in both foundation models as well as user models. Consistent with the recommendations provided in Section 4.4, this could include cultural and social biases inherent in the training data and the information conveyed by or about the user. Training data is a fundamental area for additional research. Models learn from information that is provided to them, but not everything that represents reality is written down. The user can be influenced (intentionally or unintentionally) by the results that the GenAI system produces. As a result, to make sure the user sees a result that is accurate, additional research could focus on elucidation and definition of data such as common knowledge: things that users generally know that are unlikely to be written down in a format that a model could see. The trustworthiness of the information should also be taken into account and indicated to the user. This is a key area, because of the risk of providing the user results that do not represent reality. For example, in the medical field, researchers may publish negative findings or contraindications, but not positive or neutral results. These types of research gaps or imbalances could lead into biases within data that could lead to weighting toward negative responses, which do not represent the full range of human understanding. To enable models to provide more balanced results, the research community could also consider new publication approaches that emphasize a broader range of results.

### 4.3.2. Recommendations for long-term (the next ten years) research programs

### *4.3.2.a. Cognitive models for information access*

The research community can investigate at a deep level the ability to implement cognitive models for information access. While there is the potential for significant time savings, there are also significant implications for privacy and security related to a digital twin.

### *4.3.2.b. Robust modeling of the user's state of knowledge*

Technology can focus on creating user models that can trace the user's state of knowledge in a robust way, identifying preferences and gaps and how to fill those gaps. Future GenAI-driven AI systems could take proactive steps related to filling user needs based on their knowledge of the user and their interests, in some cases preemptively providing information built from diverse sources, without the user having to enter some input. The research community could make strides toward how to plan and solve complex research tasks, driving toward the user's goal state.

*4.3.2.c. Controllable information access*

Considerations should be taken to understand the need to maintain an ability for the user to choose their actions. Rather than falling into a scenario where the GenAI capability suggests something to the user and they blindly follow the top course of action or recommendation produced by the system, there should be some capability maintained for user choice. Serendipitous discovery and novelty should be built into the systems.

*4.3.2.d. Addressing digital amnesia*

As a research community we should develop an accepted methodology to measure the effects of GenAI on people's ability to find and process and critically evaluate the results. One expected effect is amplification of so called "digital amnesia" - people degrading ability and interest to retain information that can be easily retrieved. As GenAI provides capabilities to digest and summarize and recommend and synthesize knowledge, users will need to develop new skills to evaluate the results, and the GenAI systems need to help users to use human intuition and common sense and provide background knowledge to help evaluate the answers.

## 4.4. Social ramifications

What are the consequences of developing and deploying information retrieval technology in the real world? What are the mechanisms through which information retrieval and GenAI technology bring about these consequences? And what are the corresponding risks to relevant stakeholders? In this section we provide a framework for thinking about and clarifying these questions.

Identifying risks, challenges, and opportunities of generative AI for information retrieval requires an interdisciplinary approach informed by socio-technical perspectives and co-developed with social science scholars, legal scholars, civil society representatives, and policy makers among others.

We are motivated by previous literature (Mitra et al., 2024) at surveying some of these sociotechnical implications of generative AI for information access. That work identifies several systemic consequences and corresponding mechanisms and risks of generative IR. We focus our recommendations on the "mechanisms" that represent sites of potential and concrete mitigation. That work identifies a list of 16 mechanisms that contribute to different corresponding consequences and risks: content pollution, the "game of telephone" effect, search engine manipulation, degrading retrieval quality, direct model access, the paradox of reuse, compute and data moat, AI persuasion, AI alignment, appropriation of data labor, bias amplification, AI exploitation & doxing, industry capture, pollution of research artifacts, resource demand & waste, and persuasive advertising.

Below, we indicate short-term and long-term challenges for two of these mechanisms, illustrating how we can develop research agendas corresponding to each of these mechanisms. We suspect that some of those mechanisms will not admit to this measure-and-mitigate process, creating a set of preliminary challenges.

We also describe important implications for research into socio-technical challenges in the Evaluation Campaigns section below.

### 4.4.1. Recommendations for short-term (the next five years) research programs

*4.4.1.a. The "game of telephone" effect mechanism*

This mechanism addresses the tendency for generative models to produce information that is incorrect, putting the burden of interpretation on the user: "hallucinations" is a well-known instance of this mechanism. There are broadly two complementary research directions that require attention in the short-term. (1) We need to study and develop quantitative measures for how often LLMs misrepresent information from retrieved documents in their responses, either by providing incorrect information or providing information without context that leads to misinterpretations. Correspondingly, we need to also study the impact of the incorrectness on users. (2) We also need HCI research to understand how information should be presented in a conversational context that allows for and encourages searchers to appropriately inspect and verify the presented information. For example, we should study the effectiveness of providing references in the generated responses. For e.g.,

- Do searchers pay attention to them?
- Do they actually click-through the references to verify presented information?
- Can the number of references actually mislead the searcher to trust the presented information when that's not justified?

In the same vein, we need to explore newer ways for information presentation towards empowering searchers to more effectively verify the information they are presented with. Answering these important questions will require substantial effort devoted to user studies, with the attendant funding needs that come with supporting such work.

*4.4.1.b. The AI persuasion mechanism*

This mechanism considers the possibility that generative systems may inadvertently (or intentionally) persuade users on opinions or perspectives that they do not already hold – for example, by masquerading as a trustworthy source or appealing to learned biases of the user. Correspondingly, we need research to identify and understand the mechanisms and potential mitigations. Research is also necessary to understand how persuasion capabilities interact with the incentives structures that information access systems may optimize towards, such as monetization. Recent experiences with the negative impacts of human-generated misinformation add urgency to such work, since AI will become even more pervasive with the capability of scaling without limits (so-called "Dark LLMs"). Finally, research is necessary to identify possible safeguard mechanisms and develop participatory processes and mechanisms for civil society and external scholars to investigate and audit these phenomena.

*4.4.1.c. Cost and footprint*

The computational demands of generative AI models pose challenges in community involvement in research. Historically, AI models have been small and efficient enough to run on commodity hardware, while today's large AI are effectively too inefficient and computationally demanding to do so. This has implications in the availability of these models for use by certain groups, including small organizations, students, and many underdeveloped and developing countries, which only exacerbates inequality in society given the growing role AI now plays. Advances in the efficiency of these models have the potential to democratize the technology, and more work in this space is needed. The environmental implications of the high computational of both generative AI and Neural IR are significant. Improvements to training and inference efficiency of these models – as well as the intersection between them – can help mitigate these concerns.

#### 4.4.2. Recommendations for long-term (the next ten years) research programs

*4.4.2.a. The "game of telephone" effect*

Future research questions for the "game of telephone" effect should explore how conversational agents (and LLMs in general) can be employed to intervene and encourage users to reflect on and critique the information they are presented with by posing appropriate questions, e.g., serving as a "devil's advocate". These "critical literacy" agents may also find application in mediating dialog between searchers for knowledge production and consensus building.

*4.4.2.b. AI persuasion mechanism*

For the AI persuasion mechanism, we need to develop frameworks to reflect on questions on how conversational agents should ethically interact with people, who they represent and how to be transparent about that, and what higher level objectives they are optimized for in these contexts.

### 4.5. Personalization

Improving current GenAI technology through personalization is undoubtedly an important research area with significant impact on society. Here we discuss new problems and challenges in the intersection of personalization, IR, and GenAI, which can have major applications in developing the next generation of recommender systems, question answering, writing assistants, and intelligent agents. We iterate over challenges and opportunities in developing digital twins (or *"digital shadows"*), which are capable of representing new models *per-user*, based on users' historical interactions. Such models can be used for personalized retrieval or to generate writing in a user's style. (They can also be used for evaluation as outlined earlier at *4.1.2.a*.) We argue that systems for recommendation / retrieval will gradually push the boundaries of retrieving existing content to synthesizing new content. While traditional IR can return what exists, future IR systems will generate what you want even if it doesn't exist (or a process for making it). While LLMs already synthesize content to some extent, we are specifically interested in content synthesis that is (a) more personalized to individual users than what is currently possible; and (b) able to synthesize types of content that go well beyond what is currently possible with LLMs, ranging from mixed media to (designs of) physical objects. Future personalized recommender systems will not just recommend items but might persuade a user that a specific recommendation is the best (e.g. via a conversational interface). We also highlight that personalized interfaces need to become proactive and adaptable to user needs, expectations, and preferences. We acknowledge that researchers in this area must be aware of *privacy* considerations associated with using user's data. Collecting reusable personalized datasets that also preserves user privacy is quite important in advancing personalization in generative AI models.

### 4.5.1. Recommendations for short-term (the next five years) research programs

*4.5.1.a. Personalized intelligent assistants in operating systems*
As we build stronger generative AI systems that can leverage user's personal data, preferences, and past history, we envision the development of personalized intelligent systems within operating systems that can not only execute actions, but can also retrieve and recommend files, directories, and applications. These systems should gain user trust given the information they need access to.

*4.5.1.b. Personalized dashboard for controllable digital twins*
We acknowledge the importance of providing control to the user to decide what information to be used in the development of their digital twin. This calls for the development of customizable dashboards that allow users to select what information is most relevant to the retrieval task or problem solving. Can we generate interfaces that are adaptable to users in their context and allow the user to be intentional about what information is included in the retrieval and/or generative task? Ultimately, there is a large range of possible interfaces that combine elements from "traditional" IR systems with generative components (e.g. mixed retrieval and conversational interfaces).

*4.5.1.c. Online learning for personalized intelligent assistants*
The twin must "move with the person" as they conduct their digital life. Since recent experience is often necessary for full contextualization, updates should be immediate. This calls for the development of new online learning and machine teaching algorithms for keeping personalized intelligent assistants up-to-date.

*4.5.1.d. Personalized question answering systems*
Despite their limitations in hallucination and lack of up-to-date knowledge, generative AI systems have shown great promise in answering user's questions. However, users have different background knowledge: for the same question, a user may need extended background information, while another user may need a brief and to-the-point answer. Developing resources that enable personalized question answering research and developing models that preserve the privacy of users together with providing personalized experience are among major research directions in this space.

*4.5.1.e. Modeling the trade-off between privacy and personalization*
Developing effective personalized generative AI systems requires training on and/or consuming user's personal data, which can potentially harm user's privacy. We acknowledge that this is an important and sensitive topic and should be considered seriously when designing personalized systems. Modeling the trade-off between the personalization effectiveness and privacy preservation would smooth the path towards building trustworthy personalized systems.

### 4.5.2. Recommendations for long-term (the next ten years) research programs

*4.5.2.a. Exploring various approaches for developing digital twins*
Even though immediate actions can be imagined in this space, developing effective, robust, and controllable digital twins requires significant long-term research investments. We must explore in what situations and tasks we need to use a non-personalized intelligent assistant and in what situations we must use the user's digital twin. Exploring the most efficient way to create different models tailored to specific contexts and integrate them seamlessly for holistic personalization.

*4.5.2.b. Personalized content synthesis*

Using generative AI technologies we sometimes move from retrieving information items to generating content. Such content generation approaches can better address user's information needs through synthesizing new content in a personalized way, building upon existing work on personalized search, recommendation, intelligent agents, and text generation. Preserving the privacy of users should be considered in this research.

*4.5.2.c. Persuasive information access*

Can generative AI be used to build *persuasive* recommenders that (e.g., through a conversational interface) guide users toward certain items or towards achieving personal goals? To achieve this, recommender systems should provide explanations and reasoning. These explanations are not just why the system generated a recommendation list, but why and how the user can benefit from the recommendation.

*4.5.2.d. Personalized result generation and presentation*

How can we enable users to interact with and manipulate their information in a visceral environment? Ultimately we envision IR systems in which information is presented in a way that is more personalized / contextualized to individual users, moving away from more "rigid" IR interfaces and closer to more "human-to-human" interactions.

*4.5.2.e. On-device digital twins*

As research in developing effective generative AI models on a small scale that can be handled and processed using devices with limited resources, such as smartphones, progresses, we can envision development of privacy-preserving digital twins on personal devices, such as smartphones. Here, developing in-device online updating of generative AI systems using personal data is a major challenge.

## 4.6. Scalability and efficiency

Current generative AI systems and the applications built upon them require great expertise and massive data, and computational resources. Focusing research effort and funding on more efficient systems that require fewer computational, data, and human resources lowers costs, increases access, and enables application development for smaller devices and a wider range of human activities.

Better exploiting emerging zero-shot capabilities would reduce or eliminate training and data costs in IR models. The paradigm of prompting LLMs with instruction or in-context examples has enabled the generalization to new tasks or domains without needing to collect and use training data ("zero-shot"). This paradigm can deliver substantial savings in terms of computational and data costs, and by extension human effort (i.e., task-specific training data need not be collected). Explorations on using LLMs for retrieval and ranking tasks in a zero-shot setting are fast emerging. However, leveraging these capabilities efficiently remains a key challenge. Zero-shot effectiveness appears to emerge at large model sizes, thereby requiring high computational costs to leverage. Meanwhile, these models can be distilled into smaller or more efficient architectures, though this comes with computational costs due to the additional distillation step. Therefore, the challenge remains on how to efficiently adapt LLMs to new corpora, contexts, and tasks.

Moreover, search engines will continue to be an important method of accessing information in standalone applications and as components in other systems due to their low storage, computational, and expertise requirements. LLMs will be used to generate new ways of representing queries, documents, and users that greatly improve accuracy in these efficient search engine architectures. However, existing search algorithms and data structures are based on patterns of human language usage that are well-understood. The use of new representations generated or motivated by LLMs is likely to disrupt the assumptions on which existing data structures and algorithms for efficient retrieval are based.

### 4.6.1. Recommendations for short-term (the next five years) research programs

*4.6.1.a. Continual learning*
One emerging opportunity in this domain is developing robust methods for continual learning of IR models and LLMs. Emergence of new knowledge and newer techniques will both motivate continual learning. As developing models from scratch is getting prohibitively expensive, we should develop techniques to re-use existing models and indexed corpus to build the next generation of models. This could come in different forms: (1) Updating existing models with new knowledge by updating their parameters, (2) Developing new architectures which adapt representations or parameters from existing models.

*4.6.1.b. Scaling laws for training IR models*
As GenAI models are getting more expensive to train, the community started to study how to train compute-optimal models: training over more tokens and training bigger models both will bring performance gain at the cost of compute, so how should we balance these? Studying these relationships enable researchers to determine optimal allocation of a fixed compute budget. While IR models are also getting much bigger, we lack study on such trade-offs in training compute-optimal IR models.

*4.6.1.c. Efficient training of LLMs through IR-based data curation*
Current generative AI models are trained on massive amounts of data. There is currently evidence that being selective in the process of training data curation can lead to stronger models than those that are less selective. This data selection process relates strongly to core ideas in information retrieval, including the efficient organization and clustering of data. Therefore, there is potential for IR techniques to benefit the training process of GenAI models. The benefits could take many potential forms. For instance, diversification techniques, which are well-studied in IR, could reduce the redundancy between training samples, thereby enabling the training procedure to be more sample-efficient. Alternatively, curricular learning methods could help gradually increase the estimated difficulty of the training samples, which may help GenAI models learn from appropriately challenging training samples as its training progresses.

*4.6.1.d. Model compression, distillation, quantization*
As the scale of future IR and RAG systems keeps increasing, methods for model compression, distillation, and quantization become more and more critical. Methods such as model pruning, low-rank approximations, memory and parameter efficient finetuning are going to be essential in managing the demands of large large-scale models. Similarly, improved quantization methods can be explored in query and document representations, corpus indexing, as well as the parameter space of RAG systems.

*4.6.1.e. Smaller LLMs with larger search engines*

LLMs capture knowledge about the world, knowledge about language usage, and instruction-following behavior that emerges at large scale. The most reliable way to increase accuracy has been to scale models ever larger, which increases the expense and obstacles to creating and deploying large language models in varied settings. We seek to separate these components into a smaller LLM that retains the language understanding, instruction-following, and other novel components of large language models while relying on a large and efficient search engine as the source of most information. This architecture helps with hallucination problems, is easy to update, and is far less computationally complex. It also focuses research attention on what produces desirable behaviors at scale and how to reproduce those behaviors at a smaller scale.

*4.6.1.f. Improving how the outputs from IR systems are integrated into LLM*

When augmenting LLMs with the documents retrieved from the search engine, the current practice is passing the top k documents from the IR system to the input to LLMs. Decreasing the amount of tokens prepended to LLM will improve the efficiency. Thus, future research should be directed towards investigating methods for compressing the outputs from IR systems, and selecting the minimal subset of documents needed to enable LLMs to generate valid outputs. We should also consider how optimizing IR outputs will interact with the development of long-context LLMs. See Section 4.2 for other ideas about integrating search and LLMs.

4.6.2. Recommendations for long-term (the next ten years) research programs

*4.6.2.a. New hardware paradigms*

We recognize emerging efforts in the long-term development of hardware and computational paradigms alternative to the current one based on the Von Neumann architecture. For example, quantum computing represents a groundbreaking shift in computational technology, utilizing principles of quantum mechanics to perform complex calculations at speeds unattainable by traditional computers, promising enhanced performance for computationally intensive tasks and superior energy efficiency over existing GPU systems. The reduction of energy consumption and environmental impact is essential to make generative AI and IR sustainable and scalable. Yet, it is largely unclear how current generative AI technology based on Transformers could be executed in the current experimental quantum computer. Similarly, biocomputing and neuromorphic computing promise reductions in energy consumption and high degree of parallelization. While these technologies are still underdeveloped and highly speculative, we believe the IR field should explore how techniques in the field can benefit from new and emerging computing paradigms.

*4.6.2.b. Reasoning engines are coming*

So far, we have largely considered efficiency from the compute and data perspectives. IR and GenAI have the potential to improve efficiency from the perspective of human efforts too. Current LLMs have basic reasoning and instruction-following capabilities that will only improve over time. These more powerful reasoning engines will support a wider range of information-seeking and information-usage tasks that enable new modes of accessing and analyzing information to facilitate more efficient human effort. For example, information systems will help people generate hypotheses, seek support for components of a hypothesis, and investigate information provenance and reasoning chains. The results will be faster and more powerful scientific discovery. For instance, LLMs and search techniques could improve mining patterns about protein interactions to help inform new hypotheses to test about new interactions – a very laborious process when done manually. The IR community thought about these issues in its early days, but largely drifted away from them in the web era. The community will now have the tools to investigate them more fully.

### *4.6.2.c. Constrained resources*

Computer science tends to assume that computational power and storage grow continuously over time. Research on more advanced capabilities within the same hardware footprint would provide environmental and societal benefits, as well as support future long-term human endeavors such as traveling to Mars, where the timeframe between new hardware deployments will be less rapid.

## 4.7. AI agents and information retrieval

We envision intelligent agents that are ubiquitous, effective, inexpensive and able to provide information, or accomplish tasks on behalf of users. Technologies like augmented reality enable multimodal real-time contextual input and provide multimodal output (voice, in-ear; vision as augmentations to images; information overlays on the scene). Physically-embodied agents will be able to accomplish physical tasks in the real world, acting as a bridge between the digital and analog world.

One core "tool" for an agentic IR system will be an LLM, which allows generation from IR outputs. If the user's query has multiple subtopics, the agent can issue multiple queries or a single query that covers all subtopics. The agent may have access to multiple tools, in which case it is not only generating appropriate queries but also choosing where to send them. To complete a task with these tools, an agent executes a series of actions, often called a "plan". Today these plans are often hard-coded, but autonomous planning is an essential component for future systems that interact with one or more machines, or agents. Once we have built such IR agents, we need to continue improving them. IR agents acting on behalf of people need feedback in order to adapt to individuals, their contexts, and their tasks. For example, an agent running on a device with a camera or a microphone might collect real-time implicit signals such as speech, facial expressions, gestures, or other actions the user is taking in the moment.

### 4.7.1. Recommendations for short-term (the next five years) research programs

#### *4.7.1.a. Planning*

Truly capable agents will need to be able to formulate novel plans to answer challenging questions. Several LLM-powered methods have been proposed for generating these plans and recovering from execution failures, However, the generated plans are often incorrect and new reliable methods must be developed. We also consider the efficiency of planning. We envision proactive agents that gather additional information that may lead to even more dynamic plans. To test the planning capabilities of agents we need benchmarks that are sufficiently challenging and require multiple actions to complete, as well as appropriate evaluation metrics that emphasize both the correctness of the final outputs and value of the intermediate steps. There is much to do in building evaluation platforms, and models should learn to use new tools as they are added as well.

*4.7.1.b. Interpreting feedback for Agentic IR system*

Related to the Feedback section above, moving beyond simple implicit feedback such as clicks will require more deeply understanding use behavior both for new sensors (e.g., cameras, microphones) and new presentation methods (e.g., free text, video, audio). This research can be grounded in familiar properties such as relevance and utility as well as new properties based on interaction with agents such as social norms. Second, as IR agents satisfy user needs through multi-step plans, determining the quality of a specific step traditionally requires attributing a final quality signal to a given intermediate step. In lieu of this, agents can benefit from instantaneous or partial feedback for specific steps in the plan. These partial signals can be specified by the system designer (e.g., "avoid tools that use too much time"), learned by the model (through attribution or real-time signals), or negotiated (when the tool itself is an agent). Lastly, as agentic systems become increasingly interactive, they may ask for clarification questions, or allow users—human or machine—to specify constraints or edit part of their plans. We will need to develop techniques to leverage these interventions to improve the effectiveness of future interactions.

*4.7.1.c. Verification of Agentic IR system outputs*

AI agents can generate texts containing hallucinations. How can agents expose or explain their behavior in a way that demonstrably enables human oversight? Future work should attempt to develop and evaluate (see *4.1.2.b* and *4.1.2.c*)  the effectiveness of methods such as

- **Attribution**: What excerpts from primary sources are best surfaced to provide evidence for generative summaries? Do these passages truly entail the summary? How should the authoritativeness of the primary sources be signaled?

- **Process:** For some types of information gathering queries, the *process* (or plan) followed by the agent is itself evidence of the answer's correctness. For example, given the question "Which system performs best on the XYZ benchmark" in the absence of a leaderboard, the only way to answer the question is for the agent to read each paper which cites the benchmark, extract the systems' scores, and report the max. How and when should agents expose this type of reasoning?

- **Critical Thinking:** Naive approaches to explanation generations often tend towards persuasive text that may mislead; perhaps, directing the LLM to emphasize critical thinking may lead to explanations that actually enable verification.

- **Graceful Degradation:** In the process of generating a response, an agent may fail for various reasons: some APIs it tried to access were unavailable, or returned unexpected results, or the answer was simply not available anywhere in the consulted sources. In these cases, the agent should (a) try to recover, e.g., by trying alternate ways to query the API, (b) consider alternative ways to fulfill the request, (c) if a+b fail, be able to reliably report its failure and not hallucinate an answer.

Since intuitions are often wrong, it's important to evaluate with real people in order to measure whether people can correctly recognize when the agent is correct and when returned answers are flawed.

### 4.7.1.d. Long-horizon agents

Instead of instant answers or fast generation that mimic previous generations of web search, the aim of long-horizon agents is to leverage agentic capabilities that can spend additional time planning, interacting with external tools, and iterating. It naturally includes complex multi-agent coordination with asynchronous collaboration on shared information tasks involving iterative and complex research tasks. The actions for long-term agents allow a much wider range of information actions versus existing RAG agentic models that are often a small number of calls and interactions. Long-term agents enable a much broader set of strategies leveraging interaction, external systems and tools. Such strategies could entail creation or specialization of new models as part of the task.

## 4.7.2. Recommendations for long-term (the next ten years) research programs

### 4.7.2.a. Mixed-initiative reasoning

IR agents are mixed-initiative systems that gather information proactively on behalf of the user, or accomplish tasks in the background proactively, without the direct initiation by the user.  While mixed initiative systems exist today, there are open questions about how to integrate multimodal results, when and how to take an action on behalf of the user, how to provide the user the ability to examine or verify an action that was taken, and how to manage the dialog with the user, given the recent language capabilities of LLMs.

### 4.7.2.b. Multi-agent coordination

Everyday objects (such as documents and calendars) will be imbued with intelligence to enhance the user experience.  For example, a document might detect that the people collaboratively writing need a reference to cite, and it might identify the references and add the citations to the text proactively.

### 4.7.2.c. Modular architecture

Future intelligent systems will be composite systems of multiple smaller bespoke models, with less reliance on LLMs (see *4.6.1.e*).  IR agents can be composed of multiple smaller models, and any architecture (knowledge bases, APIs, tools, other agents, etc.). This will democratize research and development in this area, as universities and smaller companies will not need vast resources to develop agents. The challenge will be developing smaller agents that communicate with each other, that produce transparent, explainable and trusted results. Evaluating modular systems presents a challenge as errors upstream compound downstream, and in modular collections of agents it may be difficult to find the source of mistakes.

*4.7.2.d. Evolving capabilities*

IR agents will need the ability to improve their effectiveness through additional interaction. They should be able to generate code that designs new IR models as well as new data processing techniques. They should be able to self-refine and continually improve through internal and external feedback.

## 4.8. Foundation models for information access and discovery

The recent technological advances in generative AI are poised to extend human capabilities further by changing how people are able to rapidly capture knowledge digitally, communicate intent with an AI-based device, and rely on the AI-device to perform numerical *and* semantic computation. We believe these new capabilities of generative AI as well as anticipated ongoing innovations will spur a new range of innovations where information access and generative AI techniques primarily serve as a way of assisting and augmenting the intelligence and abilities of every person – something we envision in an embodied form as a knowledge extender.

Inspired by the great success of general foundation LLMs as a general way to augment Intelligence and extend human capabilities, we envision the possibility of developing foundation models for supporting all kinds of information access applications. What do we mean precisely by a foundation model? A foundation model is meant to support many different kinds of applications (on top of it), thus *generality* is the first requirement. It is meant to be "task-agnostic" where these methods,tools, or systems can be applied to a wide range of tasks without being specifically designed for any single task.

### 4.8.1. Recommendations for short-term (the next five years) research programs

*4.8.1.a. Efficient, robust, and scalable generative retrieval and re-ranking*

Large transformer networks are shown to be able to produce a ranking of identifiers associated with information items in a given corpus. Large-scale efficiency and reliability of such methods, sometimes called differentiable search index or generative retrieval, are yet to be addressed or improved. Research in this area should focus on better optimization methods, effective identifier assignment, and end-to-end training as well as generalization to unseen information items.

*4.8.1.b. Instruction-following foundation models for information access*

We feel that in the domain of information access, the main inputs to the model should include: (1) information about a *user* (including the user's profile, specification, historical behavior, user intent), (2) *information sources* and collections (including APIs for information that can be accessed, document collections, the Web, (3) the *information access functionality (i.e., information access instruction)*, which defines the objectives and a space of actions, and (4) the *context* (such as the state of the outer world and the environment). The output should be a sequence of system actions and predicted user decisions, or behavior in general that best match to provide information for the particular user with the proposed content or decisions that are characterized by the context. Ideally, the system should also provide justifications and explanations of the reasoning behind the output it generates. This helps establish user trust and is a step toward making these systems more steerable. We also highlight the importance of *robustness* in this space, meaning that foundation models should be robust to various reformulations or paraphrasing of the same information access request.

*4.8.1.c. Generating a long sequence of actions for complex information seeking*

Instead of only targeting labels as output, the focus should be on multi-step chains of decisions and behaviors – where progress on this challenge will be able to simulate a user's behavior over longer future sequences. Building high-quality pre-training data through algorithmic synthetic data generation to make model training more efficient and effective is also an important direction to consider. In addition, the key to learning a foundational information access model will be to rely on techniques of learning by self-supervision, instruction, and demonstration.

*4.8.1.d. Complex multimodal information access*

To enable effective customization of such models, we also need to address challenges such as integrating multimodal input/output, determining how to best incorporate and represent a user's past behavior and decisions in the foundational layer, incorporating audio input, generating video on the fly, and providing accessible outputs customized for user's preference or current device type.

*4.8.1.e. User simulation for training foundation models*

Training these models requires diverse data sources (especially user behaviors in diverse contexts), significant computing power, and the ability to learn from small demonstrations. One could anticipate that the new foundation model will start with the behavior data from a mixture of limited real users and a large number of simulated users, and simulating user behaviors through user simulation agents is essential. Sections 4.3 and 4.5 present other ideas for how to simulate users with digital twins.

### 4.8.2. Recommendations for long-term (the next ten years) research programs

*4.8.2.a. A real-time learner knowledge extender*

Challenges will evolve to building foundational task models and user models that reflect the diversity of behaviors and objectives and that support real-time interactions in the real world. We will be able to use these models to augment human abilities – e.g. instantaneous recall of similar past patients while collaboratively listening to a patient intake, augmented reality projection and correction of physical tasks, suggestion and synthesis of information relevant to a person's current context. In a ten-year vision, the information access foundation model is anticipated to continuously learn, self-improve, mitigate and reduce data biases, and correct inherited biases from input backbone models.

# 5. ADDITIONAL RECOMMENDATIONS FOR FUNDING AGENCIES AND THE RESEARCH COMMUNITIES

## 5.1. Recommendations for evaluation campaigns

Funding evaluation pays off in three distinct ways: (1) there are evaluation artifacts such as test collections that can be used to measure systems; (2) those artifacts are built on a task concept, an understanding of what the user is trying to do in the large, that itself can be studied; and (3) the artifacts persist over a long period of time, supporting research well beyond the term of the funded program. Refer to the TREC economic impact study from 2010 (Rowe et al., 2010).

One area which information retrieval in particular holds in high regard (and in which the information retrieval community has made important advances) is evaluation. Pushing the state-of-the-art on shared reproducible benchmarks and test collections has always been important for advancing IR – and all other research areas touched on by this report. Much of the progress toward shifting the state-of-the-art across the information access community has been enabled by large-scale evaluation campaigns that challenge the community to work on a common problem and provide broadly agreed upon measures for reliably comparing system effectiveness.

These benchmarks are now common in several styles:

1. Challenges often affiliated with a conference or workshop. For example, the RecSysChallenge at the ACM RecSys conference (*Recommender Systems Challenge*, 2021), and the prototype LLMJudge Challenge run by the LLM4Eval workshop at SIGIR 2024 (*LLMJudge Challenge*, n.d.).

2. Challenges sponsored by commercial organizations. For example, the Netflix Prize and the Alexa Prize competitions.

3. Campaigns run by organizations such as NIST's TREC (in the US), CLEF (EU), NTCIR (Japan), and FIRE (India), all of which have provided venues for numerous community-wide evaluations. We note that those campaigns are well-known and trusted in the IR community, though it is not clear that they are similarly viewed outside of that community.

As information access moves more toward interactive, GenAI-based approaches, we anticipate the need for new evaluation campaigns and new demands upon the evaluations. We discuss some overarching changes that we believe should impact most of these campaigns.

Evaluation campaigns within the information access community predominantly focus on measuring how well systems can match "gold standard" output. Some campaigns are more user-focused, attempting to measure how well systems support a task, either directly or through side-by-side preference comparisons or what is commonly called A/B testing. Those sorts of evaluations have been incredibly successful in understanding and improving systems. However, in the context of societal challenges posed by GenAI, which impact a broad range of people across various professions, backgrounds, and abilities, a narrow perspective on evaluation is no longer adequate.

### 5.1.a. Support for human-centered campaigns

Experimental evaluation frameworks that do not directly leverage human input – e.g., leveraging LLMs to determine whether an answer is biased or using digital twins as synthetic users (E.* and P.*) – continue to be valuable and will likely continue to provide some gains. However, they ultimately risk a type of confirmation bias, where the LLMs agree with LLMs, ignoring human input. Unfortunately, the cost of providing human assessments is prohibitive at the scale needed for training and measuring the impacts of the mechanisms we list. For that reason, we strongly encourage efforts to support the type of human-based evaluations necessary to reduce those risks.

### 5.1.b. Engage with stakeholders

Evaluation – indeed underlying research – must engage with all feasible stakeholders to understand the impact of GenAI and IR (at least) across those groups and the individuals within them (see Societal Ramifications). We thus strongly encourage a rethinking of evaluation campaigns to be stakeholder-engaged. It is critical that specific task-based evaluations be designed and carried out – from start to finish and even beyond – in consultation with the people who are involved with that task and with anyone who is impacted by the results.

### 5.1.c. Multidisciplinary campaigns

To better understand the social issues underlying a challenge and the societal impact of potential solutions, it is necessary to involve research communities outside of computing (in addition to the stakeholders). For example, social and behavioral scientists are skilled in studying human motivations and reactions and are often much better attuned to the impact of technology on marginalized groups. As outlined in Societal Ramifications, we recommend multi-pronged campaigns where studies of users surface challenges, technology innovations are developed, their use by people is explored to identify new challenges and opportunities, technology advances, more exploration, and so on.

### 5.1.d. Non-proprietary campaigns

We strongly encourage community-wide evaluation campaigns where resources can be made available broadly, amortizing the cost of creating judgments across the entire research community.

### 5.1.e. Hardware (efficiency) aware evaluations

Evaluation practices should be geared towards modeling a heterogeneous set of controlled but realistic hardware environments that capture different operating points and constraints (e.g., environment with no GPU resources, embedded systems with specialized GPUs, small-scale, mid-scale and large-scale GPU environments). We envision these taking the form of shared tasks where such hardware environments are standardized and controlled – mimicking the familiar concept of a test-collection in TREC, but from a hardware environment standpoint. We recognize initial inroads towards addressing this goal have been recently made. For example the ReNeuIR shared task at SIGIR 2024 featured a fixed hardware environment and a common set of efficiency measurements (Fröbe et al., 2024) – but it still lacked a holistic evaluation approach, and the modeling of a diversity of constrained hardware environments.

## 5.2. Recommendations for shared computing infrastructure and resources

There is a need for greater national and international computing facilities specifically that support research on the development and use of generative AI in information retrieval systems. The funding agencies should invest in infrastructure suitable for IR-GenAI. Traditional supercomputers are often not well suited for these tasks. It may require significant storage and GPU capacity at scale to process large-scale multimodal datasets. This requires careful engineering and collocation of compute with storage (petabyte scale), high memory (nodes with multi-terabyte of memory for in-memory dense retrieval), and GPUs optimized for a mixture of both training and inference.

### 5.2.a. Open-source software for IR-GenAI

We call for significant investment in open-source tooling (such as frameworks like DSPy (DSPy: The framework for programming—not prompting—foundation models, n.d.) to maintain and evolve both core agentic tools as well as evaluation methods.

### 5.2.b. Computing research infrastructure support

We recommend computing research infrastructure grants in this area. Funding agencies should support creating such shared resources to ensure everyone can run experiments on a common ground and improve reproducibility.

### 5.2.c. A non-profit for open science for IR-GenAI.

Related to this, we recommend the creation of a non-profit organization for developing open models, open code, and open data for foundation models with applications to information access.

## 5.3. Funding programs supporting collaborative research

IR-GenAI is an interdisciplinary research topic that involves researchers from various communities, such as information retrieval, natural language processing, human-computer interaction, machine learning, and broadly artificial intelligence. We recommend funding agencies to provide support for collaborative research as major progress will not happen without the cooperation of various communities within and outside the computing field.

In addition to AI-related areas, we recommend programs that address the joint development of IR-GenAI and specialized hardware to support it. GenAI and IR systems' efficiency is constrained by the hardware capabilities upon which they are executed. Direct interaction between researchers developing new hardware and those optimizing GenAI and IR systems would ensure that future AI & IR technologies can fully leverage the next generation of hardware, ensuring short-circuiting the path from hardware ideation to usage.

# ACKNOWLEDGMENTS

## Reviewers

CCC and the authors acknowledge the reviewers whose thoughtful comments improved the report:

- Shane Culpepper, The University of Queensland, Australia
- Laura Dietz, University of New Hampshire, United States
- Rajmohan Rajaraman, Northeastern University, United States

## U.S. National Science Foundation

The workshop was supported by the Computing Community Consortium through the U.S. National Science Foundation under Grant No. 2300842. Any opinions, findings, and conclusions or recommendations expressed in this material are those of the authors and do not necessarily reflect the views of the U.S. National Science Foundation.

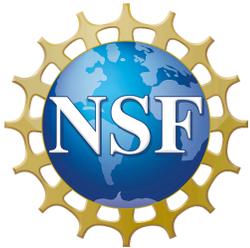

# A. APPENDICES

## A.1 Glossary

### A.1.1 Acronyms and abbreviations

- AI: Artificial Intelligence

- API: Application Programming Interface

- CCC: The Computing Community Consortium whose goal is to catalyze and empower the U.S. computing research community to pursue audacious, high-impact research. https://cra.org/ccc/

- CIIR: Center for Intelligent Information Retrieval. https://ciir.cs.umass.edu/

- CLEF: Conference and Labs of the Evaluation Forum whose goal is to promote research, innovation, and development of information access systems with an emphasis on multilingual and multimodal information with various levels of structure. https://www.clef-initiative.eu/

- CRA: Computing Research Association, a non-profit association of North American academic, governmental, and industry institutions related to computer science and engineering. https://cra.org/

- DPO: Direct Preference Optimization, an algorithm for large language model alignment.

- FIRE: Forum for Information Retrieval Evaluation. https://dl.acm.org/conference/fire

- GenAI: Generative Artificial Intelligence, models and systems that learn to generate new content, including but not limited to text, audio, image, and video.

- GPU: Graphics Processing Unit

- HCI: Human-Computer Interaction

- LLM: Large Language Model

- IR: Information Retrieval

- NLP: Natural Language Processing

- NTCIR: Japan's NII (National Institute of Informatics) Test Collection for Information Resources. https://research.nii.ac.jp/ntcir/

- RAG: Retrieval-Augmented Generation

- RLHF: Reinforcement Learning from Human Feedback

- SIGIR: ACM's Special Interest Group for Information Retrieval. Also the premiere research conference in the field. https://sigir.org

- TREC: The Text REtrieval Conference organized annually by NIST, the U.S. National Institute of Standards and Technology. https://trec.nist.gov

A.1.2. Vocabulary terms

- Digital Twin/Digital Shadow: The terms "digital twin" and "digital shadow" have their origins in work to create close digital representations of physical systems for design and testing purposes. The use of simulators by NASA in the 1960's to study and model the Apollo moon missions are considered an early example of digital twins. These concepts have been adapted and extended to a wide range of applications involving infrastructure, manufacturing and healthcare. Here, we use "digital twin" with reference to fine-grained modeling of human behavior when using IR-AI systems. While "digital twin" is currently the more common terminology, "digital shadow" is likely to be more correct in the situations contemplated herein. The former refers to a digital model that interacts with the real-world entity it is intended to represent, whereas the latter is a stand-in for a real-world entity that may not actually exist and where there is no feedback loop connection.

- IR-GenAI: The intersection of information retrieval and generative artificial intelligence research.

- Multimodal data: data with different formats, such as text, images, videos, and audios.

## A.2 CCC Workshop Participants and Report Contributors

| First Name | Last Name | Affiliation |
|---|---|---|
| Eugene | Agichstein | Emory University |
| Radhika | Agrawal | Computing Research Association |
| James | Allan | University of Massachusetts Amherst |
| Michael | Bendersky | Google DeepMind |
| Paul | Bennett | Spotify |
| Jonathan | Berant | Tel Aviv University / Google DeepMind |
| Nene | Bundu | Computing Research Association |
| Jamie | Callan | Carnegie Mellon University |
| Haw-Shiuan | Chang | UMass Amherst |
| Eunsol | Choi | UT Austin |
| Charles | Clarke | University of Waterloo |
| Arman | Cohan | Yale University |
| Nick | Craswell | Microsoft |
| Jeff | Dalton | University of Edinburgh |
| Maarten | de Rijke | University of Amsterdam |
| Fernando | Diaz | Carnegie Mellon University |
| Andrew | Drozdov | Databricks |
| Greg | Durrett | UT Austin |
| Nicola | Ferro | University of Padua |
| Grace | Hui Yang | Georgetown University |
| Petruce | Jean-Charles | Computing Research Association |
| Jean | Joyce | UMass Amherst Center for Intelligent Information Retrieval |
| Dawn | Lawrie | HLTCOE at Johns Hopkins University |
| Michael | Littman | National Science Foundation |
| Daniel | Lopresti | Lehigh University |
| Mary Lou | Maher | Computing Research Association |
| Julian | McAuley | UC San Diego |
| Timothy | McKinnon | IARPA |
| Qiaozhu | Mei | School of Information, University of Michigan |
| Bhaskar | Mitra | Microsoft Research |
| Brian | Mosley | Computing Research Association |
| Vanessa | Murdock | AWS AI/ML |
| Jian-Yun | Nie | University of Montreal |
| Negin | Rahimi | University of Massachusetts Amherst |
| Siva | Reddy | Mila / McGill |
| Mark | Sanderson | RMIT University |
| Ian | Soborrof | National Institute of Standards and Technology |
| Johanne | Trippas | RMIT University |
| Dan | Weld | Allen Institute for AI |
| Yiming | Yang | Carnegie Mellon University |
| Scott | Yih | FAIR, Meta |
| Hamed | Zamani | University of Massachusetts Amherst |
| ChengXiang | Zhai | University of Illinois at Urbana-Champaign |
| Yongfeng | Zhang | Rutgers University |
| Guido | Zuccon | The University of Queensland |